\documentclass[aps,prl,twocolumn,floatfix,preprintnumbers,showpacs]{revtex4}
\usepackage{graphicx}
\usepackage{amsmath}
\usepackage{amsfonts}
\usepackage{color}

\newcommand{\etal}{{\em et al.}}
\newcommand{\dn}[2]{d^{#1}{#2}\,}

\newcommand{\rmsubscript}[2]{{#1}_{\textrm{#2}}}
\newcommand{\lmax}{\rmsubscript{\ell}{max}}
\newcommand{\qmax}{\rmsubscript{q}{max}}
\newcommand{\rmax}{\rmsubscript{r}{max}}
\newcommand{\qscale}{\rmsubscript{q}{scale}}
\newcommand{\rscale}{\rmsubscript{r}{scale}}
\newcommand{\rmin}{\rmsubscript{r}{min}}

\newcommand{\Dgauss}{\rmsubscript{D}{gauss}}

\newcommand{\Rinv}{\rmsubscript{R}{inv}}
\newcommand{\Qinv}{\rmsubscript{Q}{inv}}
\newcommand{\qinv}{\rmsubscript{q}{inv}}
\newcommand{\Rside}{\rmsubscript{R}{side}}
\newcommand{\Rout}{\rmsubscript{R}{out}}
\newcommand{\Rlong}{\rmsubscript{R}{long}}
\newcommand{\hbarc}{\hbar c}
\newcommand{\alphaQED}{\rmsubscript{\alpha}{\tiny QED}}
\begin{document}

\preprint{\rm UCRL-JRNL-213302}

\title{
      Imaging Three Dimensional Two-particle Correlations 
      for Heavy-Ion Reaction Studies
}

%%%%%%%%%%%%%%%%%%%%%%%%%%%%%%%%%%%%%%%%%%%%%%%%%%%%%%%%%%%%%%%%%%%%%%%%%%%%%%%
\author{D.A.~Brown} 
\affiliation{Lawrence Livermore National Laboratory, Livermore California 94551}
\author{P. Danielewicz}
\affiliation{Michigan State University, East Lansing, Michigan 48824}
\author{A. Enokizono} 
\affiliation{Lawrence Livermore National Laboratory, Livermore California 94551}
\author{M. Heffner} 
\affiliation{Lawrence Livermore National Laboratory, Livermore California 94551}
\author{R. Soltz}
\affiliation{Lawrence Livermore National Laboratory, Livermore California 94551}
\author{S. Pratt}
\affiliation{Michigan State University, East Lansing, Michigan 48824}
%%%%%%%%%%%%%%%%%%%%%%%%%%%%%%%%%%%%%%%%%%%%%%%%%%%%%%%%%%%%%%%%%%%%%%%%%%%%%%%

\date{\today}

\begin{abstract}
We report an extension of the source imaging method for analyzing three-dimensional sources from three-dimensional correlations.  Our technique consists of expanding the correlation data and the underlying source function in spherical harmonics and inverting the resulting system of one-dimensional integral equations.  With this strategy, we can image the source function quickly, even with the finely binned data sets common in three-dimensional analyses.  
\end{abstract}


\maketitle

%=============================================================================
%  Main Text
%=============================================================================

%-------------------------------------------
\section{Introduction}
%-------------------------------------------

The Bose-Einstein induced correlation of particles produced in high-energy $p\bar{p}$ reactions was first observed by Goldhaber, Goldhaber, Lee and Pais~\cite{gglp} in 1960.  The connection of this effect to Hanbury-Brown/Twiss intensity interferometry followed several years later \cite{gkp_71,shuryak_80}.  Since then, two particle interferometry has become a powerful tool to study heavy-ion reactions:  it is used at all available energies from MSU-NSCL \cite{nsclreview} to RHIC \cite{rhicreview} and the Tevatron \cite{tevatron}.  The variety of particles used in these studies include protons, neutrons, mesons, intermediate mass fragments, electrons, photons, and even exotic hadrons such as $\Lambda$ and $\Xi$'s.  In all cases, our goal is to understand the space-time development of nuclear reactions.  One particularly fruitful class of this observable is multidimensional pion correlations.  Two pion correlations have been used to demonstrate the hydrodynamical scaling of the source radii \cite{mtscaling,heinz_99}, to determine the average freeze-out phase-space density \cite{wiedemann_99,bertsch_94,ferenc_99} and to characterize the final state of heavy-ion collisions \cite{heinz_99,wiedemann_99}.  We would like to extend this success to other particles, such as protons or kaons, where simple Coulomb corrections may be (or are) insufficient to study the data.  Indeed, final-state interactions (FSI) are essential ingredients \cite{pratt04} for multidimensional measurements of non-identical correlations.  For this, we extend the one dimensional source imaging technique of \cite{imag_1,imag_2,imag_3} to these multidimensional correlations.  

Source imaging allows us to cleanly separate effects due to FSI and symmetrization from effects due to the source function itself.  Because no assumptions are made about the functional form of the source function, the imaging is model independent.  This means that it is possible to image non-Gaussian sources \cite{core+halo}.  Furthermore, imaging allows us to directly compare full three-dimensional proton and pion sources, free from the distortions due to FSI.  Imaging is a useful alternate approach to the standard analysis which involves first applying a Coulomb correction then fitting the corrected data to a Gaussian.  We can extract the source {\em without} an ad-hoc Coulomb correction, and later if we wish, fit the extracted source to a Gaussian.  While this technique may be applied to other pair types, e.g. proton-proton pairs  \cite{lisa-pp}, we will concentrate on like meson pairs.  

Extracting the source function, $S_{\bf P}({\bf r'})$, for a pair of identical particles begins by noting that the Koonin-Pratt equation relates it to the experimentally measured two-particle correlation, $C_{\bf P}({\bf q'})$ (or ${\cal R}_{\bf P}({\bf q'})$, its deviation from 1) \cite{pratt_90,koonin_77}:
\begin{equation}
	{\cal R}_{\bf P}({\bf q'})\equiv C_{\bf P}({\bf q'}) -1 =
	\int d{\bf r'} \,K({\bf q'}, {\bf r'}) \, S_{\bf P} ({\bf r'}) \,.
	\label{eqn:KP3D_K}
\end{equation}
Thus, ``imaging the source'' means inverting this equation.  In \eqref{eqn:KP3D_K}, ${\bf P}={\bf p_1}+{\bf p_2}$ is the total momentum of the pair and the kernel of the integral equation is 
\begin{equation}
	K({\bf q'},{\bf r'}) = |\Phi_{\bf q'}^{(-)}({\bf r'})|^2-1.
	\label{eqn:kernel}
\end{equation}
The wavefunction, $\Phi^{(-)}$, describes the propagation of the pair from a relative separation of ${\bf r'}$ in the pair center of mass (CM) to the detector with relative momentum ${\bf q'}=\frac{1}{2}({\bf p_1'}-{\bf p_2'})$.  Here primes denote quantities in the pair CM frame. The source function itself is the probability of emitting the pair a distance of ${\bf r'}$ apart in the pair CM frame.
 
The problem of inverting Eq.~\eqref{eqn:KP3D_K} is generally ill-posed, meaning that small fluctuations in the data, even if well within statistical or systematic errors, can lead to large changes in the imaged source function.  In angle-averaged correlations, this stability problem is solved by the use of constraints, a basis spline representation of the source, and ``optimized discretization'' \cite{imag_1,imag_2,imag_3}.  By expanding the source and the correlation in spherical harmonics, we may convert the full three-dimensional problem into a series of one-dimensional problems where these tools are equally applicable.  Throughout this process, we must take care to  respect the frame dependence of the source function.  In an angle-averaged like-pair correlation, one may work  in terms of $\qinv=\sqrt{{\bf q}^2-q_0^2}$, so the analysis frame is not an issue.  In full three-dimensional analysis, we must properly carry out the transformation from the frame where the data is compiled to the frame where we image.

We now outline this paper.  First, we will set up the problem, taking into account the special issues in a full three-dimensional analysis.  This includes a discussion of how the source function and underlying emission functions are related, the frame dependence of the imaging, the coordinate system used in the imaging and a non-trivial test case.  Next, we describe the three-dimensional imaging procedure.  This process amounts to breaking the problem into a series of one-dimensional problems.  We describe this procedure and its implications, especially to the source and its representation.  We illustrate the entire imaging process using the non-trivial test case.  

%-------------------------------------------
\section{Preliminaries} 
%-------------------------------------------

In this section, we detail the connection between the correlation and source functions and the underlying emission functions.  In this process, we illuminate some of the technical issues arising from the frame dependence of the source function.  We also define the coordinates that we use in our analyses.  Following this, we detail the non-trivial model correlation we use as a test case to illustrate our analysis procedure.     

%\subsection{The Analysis Frame and Interpretation of the Source Function}

\subsection{Relating the Source Function and Emission Functions}

Before we can image, we write the Koonin-Pratt equation in a covariant form.  This is important because the imaging is simplest in the pair CM frame, yet the data is often tabulated in an entirely different frame.  We begin with the measured correlation which written as a convolution of the final state wavefunction with a relative distance distribution:
\begin{equation}\begin{array}{rl}
   C_{\bf P}({\bf q})
      &= \left.\displaystyle 
            E_1 E_2\frac{\dn{}{N_{12}}}{\dn{}{\bf p_1}\dn{}{\bf p_2}}
         \right/
         \left(\displaystyle E_1\frac{\dn{}{N_{1}}}{\dn{}{\bf p_1}}\right)
         \left(\displaystyle E_2\frac{\dn{}{N_{2}}}{\dn{}{\bf p_2}}\right)\\
      &=\displaystyle \int\dn{4}{r}\left|\Phi_{\bf q}(r)\right|^2 
         \mathcal{D}_{\bf P}(r)
\end{array}\label{eqn:bigfancyKPeqn}\end{equation}
Here the relative distance distribution, $\mathcal{D}_{\bf P}(r)$, is the probability of emitting a pair with a space-time separation $r=(r_0,{\bf r})$.  Each particle has approximately momentum ${\bf P}/2$ (in the smoothness approximation, the slight difference between using ${\bf p_1}$, ${\bf p_2}$ and ${\bf P}/2$ is ignored \cite{rhicreview}).  The relative distance distribution can be written as the convolution of emission functions for particles 1 and 2:
\begin{equation}
   \mathcal{D}_{\bf P}(r)=\int \dn{4}{R} D_1(R+r/2,{\bf P}/2) D_2(R-r/2,{\bf P}/2).
   \label{eqn:relDist}
\end{equation}
The emission functions are the normalized particle emission rates:
\begin{equation}
   D(r,{\bf p})=\left.\frac{E\dn{7}{N}}{\dn{4}{r}\dn{3}{p}}\right/
	\frac{E\dn{3}{N}}{\dn{3}{p}}.
\end{equation}
This rate could in principal be computed in models such as RQMD \cite{RQMD} or the Blast-Wave \cite{blastwave}.

To make contact with Eq. \eqref{eqn:KP3D_K}, we boost to the pair CM frame. Since Eq. \eqref{eqn:bigfancyKPeqn} is written covariantly, we may re-express it in any frame simply by replacing coordinates in the original frame with coordinates in the pair CM frame.  Further, since the the wavefunction is independent of the relative time, $r_0'$, in the pair CM frame \footnote{The time dependence appears as $e^{\pm ir_0'E_\textrm{rel}}$ factors in the two terms of a (anti-)symmetrized pair wavefunction or the one term of an unlike pair wavefunction.  For like pairs, $E_\textrm{rel}=0$, removing the time dependence.  For unlike pairs, this phase-factor is removed by the square in \eqref{eqn:kernel}.}, the time integral may be brought over to act on the relative distance distribution:
\begin{equation}
   C_{\bf P}({\bf q}) =\int\dn{3}{r'}
      \left|\Phi_{\bf q'}({\bf r'})\right|^2 
         \int \dn{}{r_0'} \mathcal{D}_{\bf P}(r)
\end{equation}
%Here we have used the fact that ${\bf P'}=0$ in the pair CM frame.  We have also added an index to the relative distance distribution to denote the momentum corresponding to the boost.  
This allows us to define the source function in the pair CM frame:
\begin{equation}
   S_{\bf P}({\bf r'})=\int \dn{}{r_0'} \mathcal{D}_{\bf P}(r).
\label{eqn:sou-def}\end{equation}
With this source, we arrive at the Koonin-Pratt equation in Eq.~\eqref{eqn:KP3D_K}.  %Written in this last form, Eq.~\eqref{eqn:KP3D_K} does not appear to be covariant.  Still, this equation is covariant as the source and differential volume element must transform together under a boost. 

%We can use Eq.~\eqref{eqn:KP3D_K} in a frame other than the pair CM frame by regarding ${\bf q'}$ and ${\bf r'}$ as variables in the frame of interest.  Explicitly, 
%\begin{equation}\begin{array}{rll}
%   {\bf q'}&=\left(\gamma (q_\|-\beta q_0), {\bf q_\perp}\right)&\\
%   {\bf r'}&=\left(\gamma (r_\|-\beta r_0), {\bf r_\perp}\right)&
%\end{array}\label{eqn:booststuff}\end{equation}
%where $\beta$ is the boost velocity, and the $\|$ and $\perp$ directions correspond to directions parallel and perpendicular to the boost velocity, respectively.  For like pairs, the result for ${\bf q'}$ in Eq. \eqref{eqn:booststuff} simplifies to ${\bf q'}=\left(q_\|/\gamma, {\bf q_\perp}\right)$ since $q_0'=0$, the particles are on mass shell, and the boost velocity to the pair CM is $\boldsymbol{\beta}={\bf P}/P_0$.

\subsection{The Coordinate System}

\begin{figure}
   \includegraphics[width=0.45\textwidth]{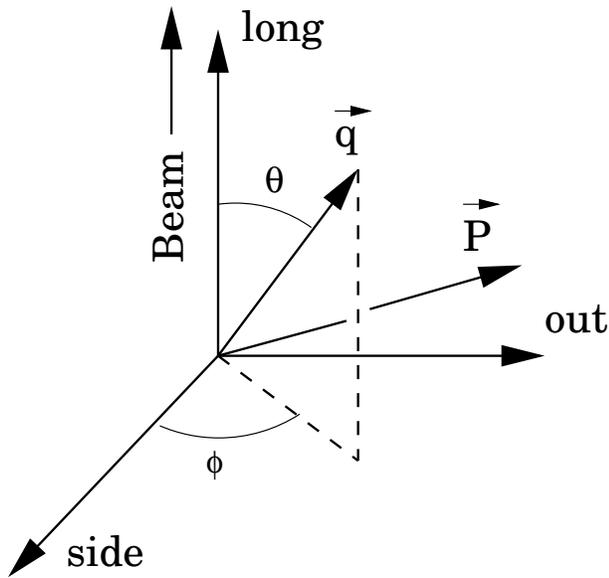}
   \caption{The Bertsch-Pratt coordinate system in ${\bf q}$-space.}
   \label{fig:coordinates}
\end{figure}

We use the Bertsch-Pratt coordinates \cite{pratt_90,bertsch_88} in both ${\bf r}$ and ${\bf q}$ spaces and we define these coordinates in the rest frame of the interacting nuclear system.  Once the directions of the unit vectors are defined, we can use the coordinate system in any frame boosted relative to the system frame.  In the Bertsch-Pratt coordinates, the longitudinal direction ${\bf \hat{e}}_L={\bf \hat{z}}$ is along the beam direction and the sideward direction ${\bf \hat{e}}_S$ is perpendicular to the longitudinal direction and the direction of the total pair momentum, i.e. ${\bf \hat{e}}_S={\bf P}/|{\bf P}|\times{\bf \hat{e}}_L={\bf \hat{x}}.$  The final direction, the outward direction, lies in the plane defined by the beam direction and ${\bf P}$.  The unit vector in the outward direction ${\bf \hat{e}}_O$ may be obtained from the other two unit vectors: ${\bf \hat{e}}_O={\bf \hat{e}}_L\times{\bf \hat{e}}_S={\bf \hat{y}}$.  Together, the three unit vectors define the right-handed coordinate system shown in Fig. \ref{fig:coordinates}.  In this figure, we also define the polar angles $(\theta,\phi)$ which we use when working in spherical coordinates.

Our use of these coordinates does not preclude performing an azimuthal HBT-type analysis.  In an azimuthal analysis, the part of the total momentum vector {\bf P} perpendicular to the longitudinal axis is further decomposed into components parallel to and perpendicular to the reaction plane.  Since the Bertsch-Pratt coordinates are defined relative to the {\bf P} direction, a correlation (or source) in these coordinates can be measured (or imaged) with respect to the reaction plane.

\subsection{Simple Test Problem}
\label{testproblem}

To test the imaging codes, we need to generate a test source function and correlation.  To do this, we first choose the single particle source $D({\bf r},t,{\bf p})$ to use in Eqs. \eqref{eqn:KP3D_K} and \eqref{eqn:sou-def}.  Then we perform the integrals in Eq. \eqref{eqn:KP3D_K} using the Correlation AfterBurner ({\tt CRAB}) code \cite{crab} to generate the correlations.  We use the symmetrized relative wavefunction, including Coulomb repulsion.  

In the lab frame, our single particle source is composed of a prompt source and a long-lived source:
\begin{equation}\begin{array}{rl}
   D({\bf r},t,{\bf p})\propto &\displaystyle\left(f\delta(t)+(1-f)\theta(t)\exp \left(-t/\tau\right)\right)\\
   & \displaystyle\times\exp \left(-\frac{{{\bf p}\;}^2}{2m_\pi T}\right) \Dgauss({\bf r}).
   \label{eqn:emission_func}
\end{array}\end{equation}
Here, the Gaussian given by
\begin{equation}
   \Dgauss({\bf r})\propto
   \exp\left(-\frac{x^2}{2R_x^2}-\frac{y^2}{2R_y^2}-\frac{z^2}{2R_z^2}\right).
\end{equation}
Our test single particle source is radially symmetric with $R_x=R_y=R_z=4$~fm, with a life-time of $\tau=20$~fm/c (comparable to the $\omega$ lifetime of 23~fm/c and the system lifetime in other models \cite{bertsch_88}).  The fraction of pions emitted from the prompt versus long-lived source is controlled by $f$ and is set to $f=0.65$, corresponding roughly to the number of pions created from all resonance decays in reaction simulations~\cite{RQMD,sullivan}.  We set the source temperature to 50~MeV.
\begin{figure}
   \includegraphics[width=3.25in]{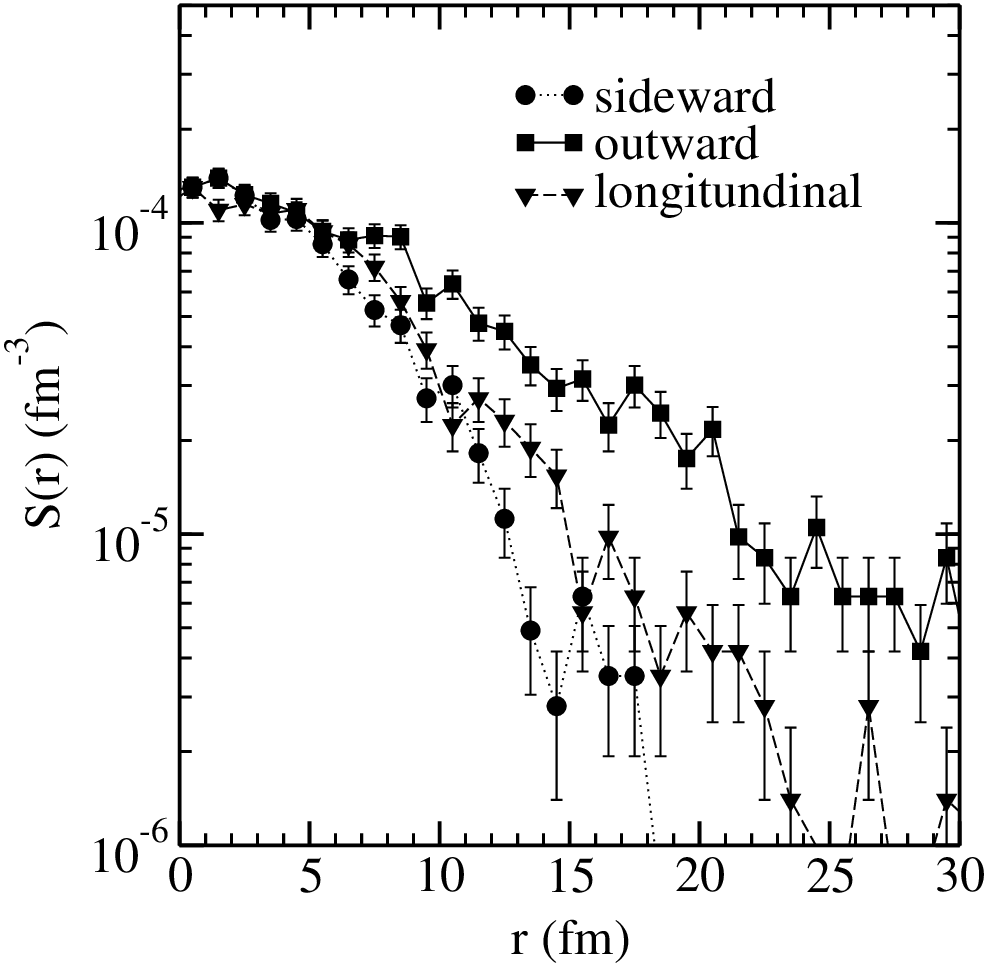}
   \caption{The model source function constructed from the model emission functions in Eq.~\eqref{eqn:emission_func}.}
   \label{fig:model_sou}
\end{figure}

To produce the source function in Fig. \ref{fig:model_sou}, we first generated a set of phase-space points sampled from the single particle source in Eq. \eqref{eqn:emission_func}.  We then perform the integrals in Eqs. \eqref{eqn:relDist} and \eqref{eqn:sou-def} using this set of phase-space points.  Since we did not have pairs with exactly the same momentum, we kept pairs with relative momentum less then 60 MeV/c when computing the source function \cite{panitkin_99_1}.  We investigated whether this choice of affects the source function, but found no noticeable dependence.  Because of the low temperature of the single particle source, we include all simulated pairs (925260 altogether) in the calculation.  In Fig. \ref{fig:model_sou}, we see that the resulting source function is both non-Gaussian and non-spherical as shown in the outward and longitudinal curves.  

\begin{figure}
	\includegraphics[width=0.45\textwidth]{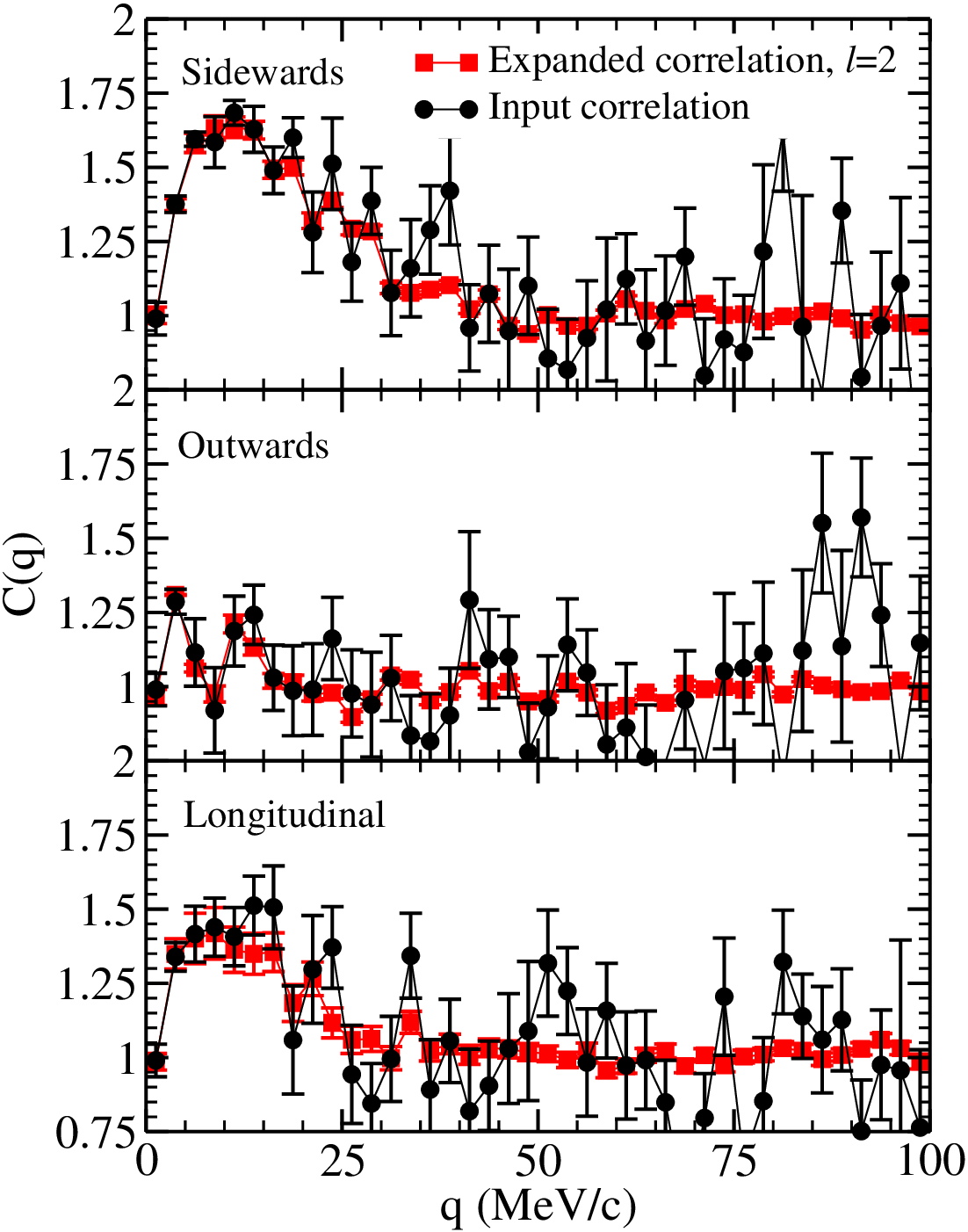}
	\caption{The input model correlation compared with the correlation expanded in spherical harmonics, stopping at order $\lmax=2$.  In the panels, we show 5 MeV wide slices cut along each axis.}
	\label{fig:model_corr}
\end{figure}

Our test correlation is shown in Fig.~\ref{fig:model_corr} in the black points.  We have chosen a much finer binning than is usually done in correlation analyses to gain resolution in the expanded correlation as we describe in the following section.  One can clearly see that this test data has {\em not} been Coulomb corrected in any way and that the correlation is not spherically symmetric.  Furthermore, one can see that the tail in the outward direction has pushed most of the correlation in the outwards direction into the Coulomb hole. 

%-------------------------------------------
\section{The Imaging Procedure}
%-------------------------------------------

In an experiment with resolution $\Delta q=10$~MeV/c, there are 40 bins in each direction if one measures to relative momentum difference $q\sim 200$~MeV/c.  This means that there are $40^3$=$64\times 10^3$ elements in the data set.  If we use a Cartesian grid for representing the source with 10 bins in each direction, then we will need to perform $64\times 10^3 \times 10^3 = 64\times 10^6$ six-dimensional integrals just to create the inversion matrix.  Clearly it is to our benefit to work to reduce the size of the problem.  Fortunately, the full three-dimensional imaging problem can be broken up into a series of 1d inversions which are much easier to handle numerically.  %All told, it will require evaluating roughly $2.3\times 10^7$ three-dimensional integrals to expand the correlations in spherical harmonics and perhaps another 3600 two-dimensional integrals to image the source, clearly a dramatic reduction in computational burden.

In this section, we first describe how we convert the full three-dimensional problem into a series of one-dimensional inversion problems by expanding the correlation and source in spherical harmonics and the kernel in Legendre polynomials.  Next, we explain how we expand the data itself in this basis.  Following this, we explain our procedure for setting cut-off value $\qmax$ for all the radial integrals and the cut-off order $\lmax$ in our expansions.  Fourth, we detail the representation of the source function and we go into some detail how the kernel determines the location of the knots in the basis spline expansion of the source.  The knot locations follow from the Sampling Theorem in Fourier theory and require some discussion.  Finally, we describe the imaging itself.

\subsection{Converting the 3d problem into a series of 1d problems}

We begin by expanding the source and correlation in spherical harmonics:
\begin{equation}
	S({\bf r}) = \sqrt{4\pi} \sum_{\ell=0}^{\lmax} \sum_{m=-\ell}^{\ell}
		Y_{\ell m}(\hat{\bf r}) S_{\ell m}(r)
	\label{sou_1d}
\end{equation}
and
\begin{equation}
	{\cal R}({\bf q}) = \sqrt{4\pi} \sum_{\ell=0}^{\lmax} 
		\sum_{m=-\ell}^{\ell}
		Y_{\ell m}(\hat{\bf q}) {\cal R}_{\ell m}(q).
	\label{corr_1d}
\end{equation}
We also expand the kernel in Legendre polynomials in $\mu=\cos(\theta_{{\bf q},{\bf r}})$:
\begin{equation}
	K({\bf q}, {\bf r}) = \sum_{\ell=0}^{\lmax} 
		P_{\ell}(\mu) K_{\ell}(q,r).
	\label{kern_1d}
\end{equation}
After inserting these into the Koonin-Pratt equation, we simplify the result with the Spherical Harmonic Addition Theorem.  Doing so, the Koonin-Pratt equation becomes a system of linearly independent integral equations:
\begin{equation}
	{\cal R}_{\ell m}(q) = 4\pi\int_0^\infty \dn{}{r} r^2 K_\ell(q,r) S_{\ell m}(r).
    \label{KP_expanded}
\end{equation}
Instead of attempting a full three-dimensional inversion, we will attack this system of 1d inversions using the imaging techniques developed in Refs.~\cite{imag_1,imag_2,imag_3}.  

Since we focus on like-meson pairs in this paper, we write the kernels that we need.  For noninteracting spin-0 bosons the kernel $K_\lambda(q,r)$ is
\begin{equation}
   K_\lambda(q',r')=
      \left\{
         \begin{array}{lr}
            (-1)^{\lambda/2}j_\lambda(2qr) &\mbox{for $\lambda$ even} \\
            0 &\mbox{otherwise.}
         \end{array}
      \right.
      \label{NIbosonkernel}
\end{equation}
If the bosons do interact, the kernel becomes a bit more complicated:
\begin{equation}
   K_\lambda(q,r)=\sum_{\ell \ell'} y_{\ell}(r) y_{\ell'}^*(r)
      \left(\begin{array}{ccc}
         \lambda & \ell & \ell' \\
         0 & 0& 0
      \end{array}\right)^2 - \delta_{\lambda 0}
\end{equation}
We typically truncate the sum over $\ell$'s at several times $\qmax\rmax/\hbarc$, where $\qmax$ and $\rmax$ are the maximum values we will need for a particular data set.  Here, the object in the parentheses is a Wigner $3-j$ symbol and the $y_\ell$ are the symmetrized solutions of the radial Klein-Gordon equation:
\begin{equation}
	y_\ell(r) =\left\{
	\begin{array}{ll}
		\sqrt{2}(-1)^{\ell/2} (2\ell+1)e^{i\sigma_\ell} F_\ell(\eta;\rho)/\rho & \text{even $\ell$}\\
		0 & \text{odd $\ell$}
	\end{array}
	\right.
\end{equation}
where $\sigma_\ell$ is the Coulomb phase, $F_\ell(\eta;\rho)$ is the Coulomb radial wave as defined by Messiah \cite{messiah}, $\rho = qr/\hbarc$ and $\eta = \alphaQED\sqrt{q^2+m^2}/q$.  While we must include the Coulomb interaction between the pair, we may neglect the short-range meson exchange potential.  

\subsection{Converting the 3d data into 1d data}

Experimentally measured correlations are inevitably binned in ${\bf q}$ and usually in fixed size Cartesian bins.  In other words, rather than measuring a continuous correlation, ${\cal R}({\bf q})$, one measures the average value of the correlation over a bin: 
\begin{equation}
   {\cal R}_{(i)}
      ={\cal R}({\bf q}_{(i)})
      \equiv\frac{1}{V_{(i)}}\int_{V_{(i)}} \dn{3}{q}{\cal R}({\bf q}).
      \label{binnified_c2}
\end{equation}
Here the bin index $(i)$ is shorthand for the set of three indices $\left\{i_S,i_O,i_L\right\}$ and the volume of the $(i)^{th}$ bin as $V_{(i)}$.   For fixed sized binning, $V_{(i)}$ is constant and given by $V_{(i)}=\Delta q_S\Delta q_O\Delta q_L$.  If we introduce a ``window function,'' $w_{(i)}({\bf q})$, where 
\begin{equation}
w_{(i)}({\bf q}) =\left\{ \begin{array}{rl}
    1 & \textrm{if ${\bf q}$ in $V_{(i)}$} \\
    0 & \textrm{otherwise,}
\end{array}\right.
\end{equation}
we can write
\begin{equation}
	{\cal R}({\bf q}) = \sum_{(i)}{\cal R}_{(i)}w_{(i)}({\bf q}).
    \label{binnified_c2_cute}
\end{equation}

For like pairs, we can reduce the number of data points by a factor of two if we notice that the correlation data must be invariant under particle label switching.  In relative coordinates, this is equivalent to:
\begin{equation}
   {\cal R}({\bf q})= {\cal R}(-{\bf q}).
   \label{parity_inversion1}
\end{equation}
If we align the edges of the bins along the coordinate axes, then we can reflect half of the data back on itself, simultaneously reducing the size of the data set by half and doubling the statistics on the other half.  For the purpose of this paper, we will assume all ${\bf q}$-bins with $q_S<0$ have been properly reflected into the $q_S>0$ bins.  

We can now convert a binned correlation in Cartesian coordinates to a series of binned correlations in spherical coordinates.  We invert Eq. \eqref{corr_1d} and average over radial bins:
\begin{equation}
	{\cal R}_{\ell m j}=\frac{1}{\Delta q_j}\int_{q_j-\Delta q_j/2}^{q_j+\Delta q_j/2}\dn{}{q}\int_{4\pi}\dn{}{\Omega_{\hat{\bf q}}}Y^*_{\ell m}(\hat{q}){\cal R}({\bf q}).
\end{equation}
Inserting this into \eqref{binnified_c2_cute}, we find 
\begin{equation}
	{\cal R}_{\ell m j} = \sum_{(i)} {\cal M}_{\ell m j (i)}{\cal R}_{(i)}
\end{equation}
where the transform matrix ${\cal M}$ is 
\begin{equation}
    {\cal M}_{\ell m j (i)} = \frac{1}{\Delta q_j}\int_{q_j-\Delta q_j/2}^{q_j+\Delta q_j/2}\dn{}{q}\int_{4\pi}\dn{}{\Omega_{\hat{\bf q}}}Y^*_{\ell m}(\hat{q})w_{(i)}({\bf q}).
\end{equation}

Now, the data is a collection of binned one-dimensional correlations suitable for imaging.  This procedure has the side effect of introducing radial bin-to-bin correlations since the Cartesian bins are ``non-local'' in spherical coordinates.  In fact, the uncertainties are non-trivially distributed among all the $\ell m$'s since the bins in the data are re-used to compute each term.  Maintaining these correlations would require us to treat all terms simultaneously, resulting in a numerically intractable problem.  To make progress, we ignore the cross-$\ell m$ correlations, but maintain the bin-to-bin correlations.  They are quantified in the covariance matrix of the radial correlation:
\begin{equation}
    \Delta^2{\cal R}_{\ell m jj'} = \sum_{(i)}{\cal M}_{\ell m j (i)}{\cal M}_{\ell m j' (i)}(\Delta{\cal R}_{(i)})^2.
\end{equation}
In other words, the uncertainties in the correlation data are non-trivially distributed among the radial bins in the expanded correlation.  

\begin{figure}
	\includegraphics[width=0.45\textwidth]{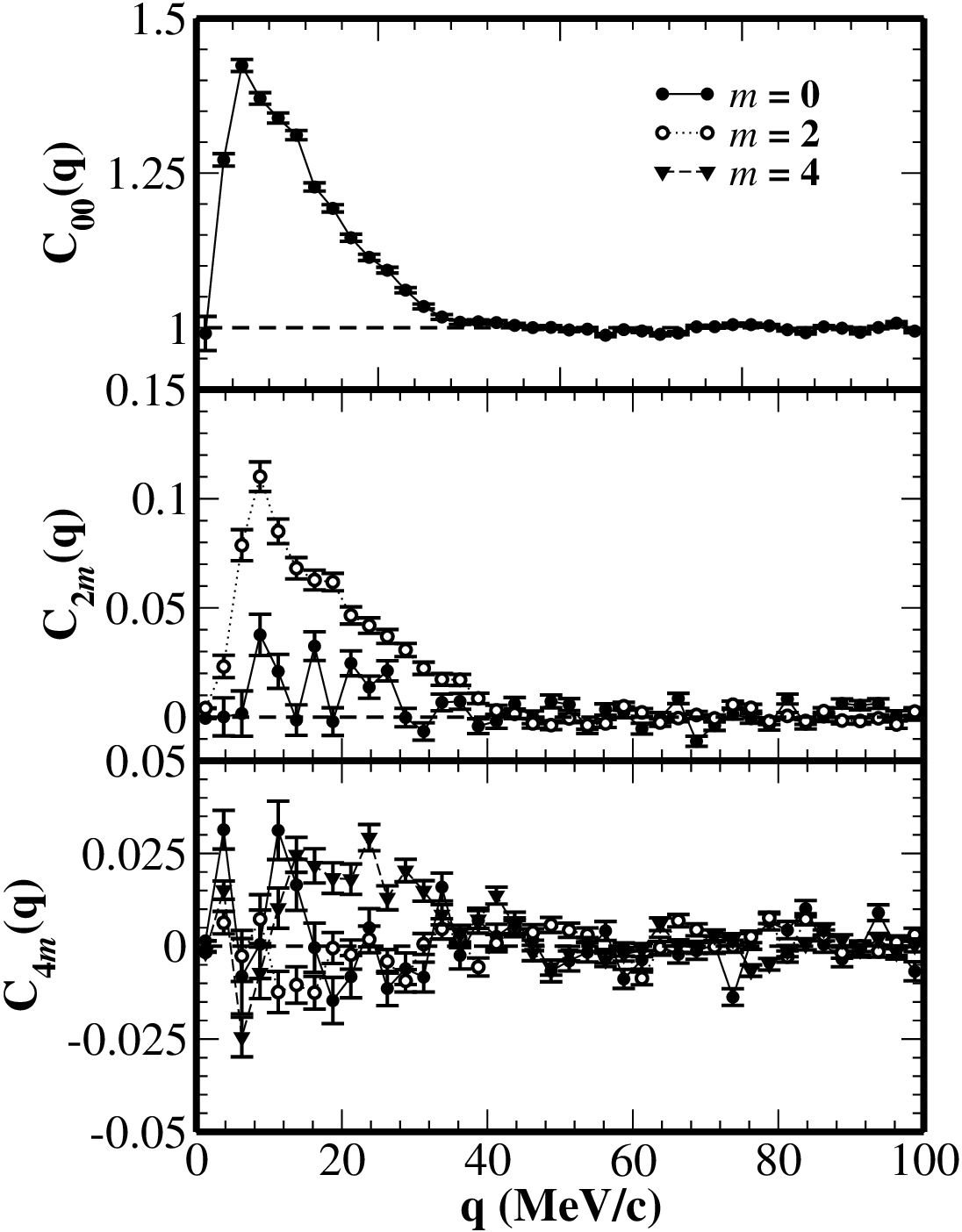}
	\caption{Terms in the spherical harmonic expansion of the model correlation.  Since this is a like pair correlation, all odd $m$ terms are zero.  Because of the symmetry built into our model correlation, all terms have only real parts.}
	\label{fig:corr_expanded}
\end{figure}

We apply this scheme to the test correlation.  In Fig. \ref{fig:corr_expanded} we show the individual terms in the expansion.  We only show $\ell \le 4$ terms, although we generated all terms up to $\ell = 10$ from the test data.    
In Fig.~\ref{fig:corr_expanded}, we recognize two important features.  First, we see that the Coulomb hole is only noticeable in the $\ell m=00$ term.  The higher $\ell$ terms also have a dip at the origin but this is a result of the analyticity of each term of the correlation rather than a Coulomb induced effect.  The second thing we notice is that the statistical fluctuations in the expanded data are greatly reduced.  This results from the extensive angular averaging that occurs in the low $\ell$ terms.  Thus, by picking a seemingly unreasonable binning in ${\bf q}$ in the original correlation, we can achieve a much higher resolution in the individual terms of the expanded correlation.  As we go up in $\ell$, the fluctuations become more intense as we begin to resolve the angular structure of the starting bins.  This will require us to cut off our spherical harmonic expansion at a relatively low $\ell$.  

In Fig.~\ref{fig:model_corr} we show the (reassembled) expanded correlation in the squares alongside the original in the circles.  It is clear that truncating the spherical harmonic expansion at a relatively low $\ell$ (2 in this case) has not distorted the correlation and has resulted in significant smoothing of the correlation.  Some of the smoothing may actually be removing information from the correlation, but for the most part the smoothing  removes angular noise.  

We must truncate our source expansion at some angular scale $\ell_{max}$.  Care must be taken in choosing this $\lmax$: if $\lmax$ is too low, then the higher angular scale structure in the source will be lost.  Actually, the problem is worse than this -- this higher angular structure can be mistaken for noise in the data \cite{scales_tutorial,cobe_1,cobe_2}, interfering with the angular components we otherwise seek.  On the other hand, if $\lmax$ is too large, other problems may arise.  A large $\lmax$ may be large enough to sample the high frequency angular structure imposed by the Cartesian binning at large $q$. Also, since the number of pairs often drop at $q$ increases, statistical fluctuations in the high-$q$ bins may result in {\em angular} aliasing.  For a discussion of aliasing, see either Ref.~\cite{imag_3} or~\cite{NumRecipes}.  In practice, we can set both $\lmax$ and $\qmax$ simply by visually inspecting the terms in the correlation expansion.  In the next subsection we will introduce a coarser scheme for setting $\lmax$ and $\qmax$ that is nevertheless interesting because of its relation to the space-averaged phase-space density.

We conclude this subsection by describing the physical meaning of the terms in the expansion: 
\begin{itemize}
	\item[$\ell=0$:]  Angle integrated shape.  The $\ell m = 0 0$ term {\em is} the $q_{\rm inv}$ correlation if one is analyzing a like pair correlation.  This term is sensitive mainly to $\Rinv$.
	\item[$\ell=1$:]  Dipole distortion.  The longitudinal and outwards directions are sensitive to any emission time offset (i.e. the ``Lednicky offset'' \cite{lednicky}).  If the temporal offset occurs in the longitudinally co-moving frame, the offset will show up in the $\ell m=11$ term.  If the offset occurs in the colliding nuclei's frame, this offset will show up in both the $\ell m=10$ and $\ell m=11$ terms.  These terms are  absent for like pairs.
	\item[$\ell=2$:] Quadrupole distortion.  These terms give the main shape of the correlation so are sensitive to the deviation of $\Rside$, $\Rout$, and $\Rlong$ from $\Rinv$.  The $\ell m = 20 $ term gives $\Rlong$ while linear combinations of the $\ell m = 00, 20,$ and $22$ terms give $\Rside$ and $\Rout$.
	\item[$\ell=3$:] Octapole or triaxial distortion.  These terms are responsible for making the source look something like a ``boomerang'' and are absent for like pairs. 
	\item[$\ell=4$:] Hexadecapole distortion.  These terms are responsible for making the source more ``box-like.''
\end{itemize}

\subsection{Integrals of the data}

In this section, we introduce two integral measures of the correlation data: the ``bump volume'' and the spectral power.  These measures can be used to 
estimate both $\lmax$ and $\qmax$, but are perhaps more useful as a gross characterization of the structure of the full three-dimensional correlation.  

\begin{figure}
	\includegraphics[width=0.45\textwidth]{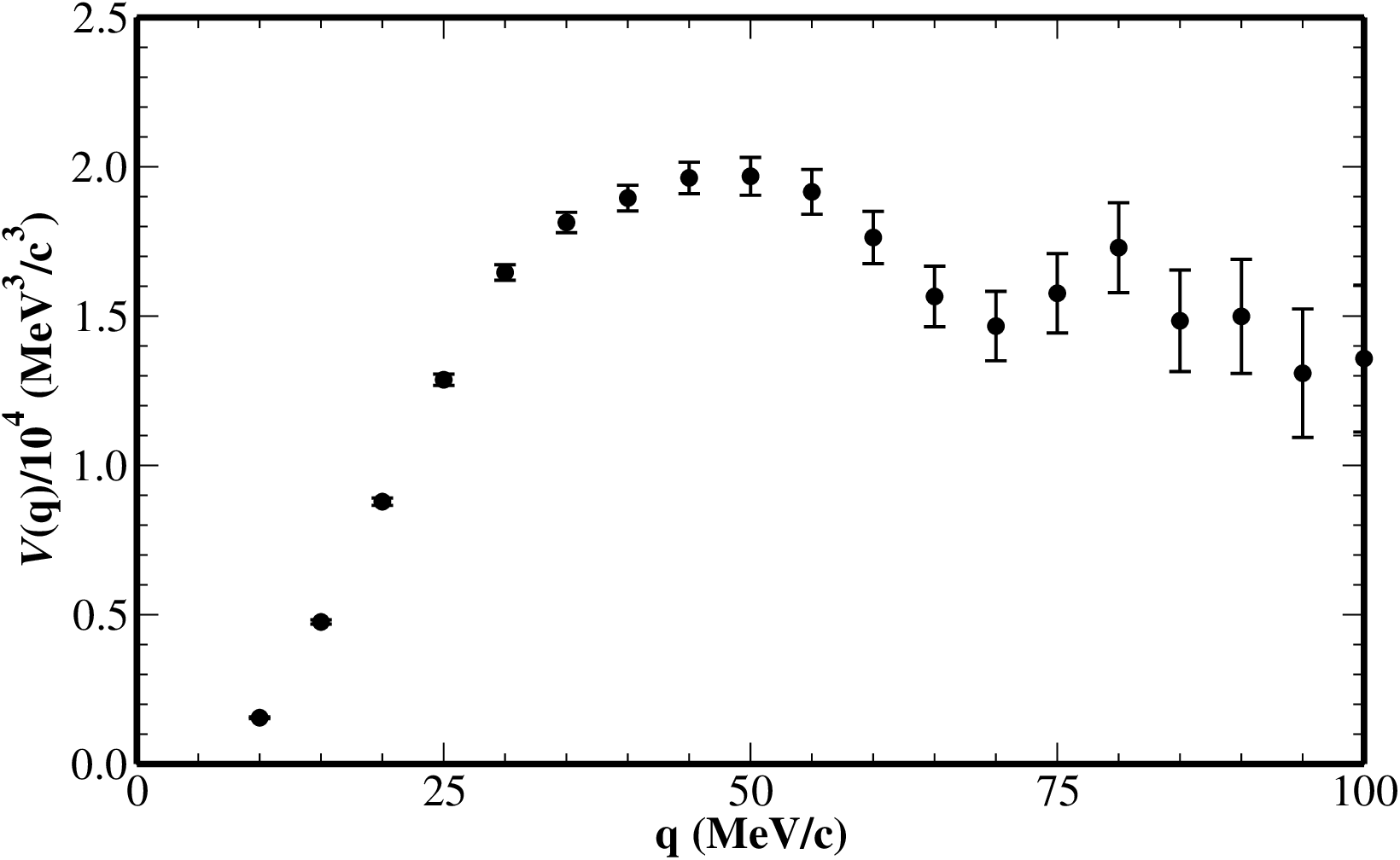}
	\caption{``Bump volume'' of the model correlation.  }
	\label{fig:bump_integral}
\end{figure}

The ``bump volume'' is simply the volume integral of the correlation:
\begin{equation}
     V(\qmax) = 4\pi \int_0^{\qmax} dq q^2 {\cal R}_{00}(q)
\end{equation}
This corresponds to the ``bump volume'' \cite{bumpvol} and is directly related to the space averaged phase-space density \cite{imag_1}, $\left<f({\bf p})\right>$, for identical non-interacting pairs through 
\begin{equation}
    \left<f({\bf p})\right> = \pm \frac{8}{(2s+1)m}\frac{Ed^3N}{d^3p} {\cal V}(\qmax).
\end{equation}
This relation is a direct consequence of Eq. (17) in Ref. \cite{imag_1} and the sum rule of the correlation described in Ref. \cite{sumrule}.  Unfortunately, as noted in Ref. \cite{sumrulesucks}, this sum rule can not be applied reliably to data in the presence of final-state interactions.  Nevertheless, we have plotted this volume in  Fig.~\ref{fig:bump_integral} for the test problem.
We note that the integral is essentially flat past $q=35$ MeV/c, indicating there is no information accessible in the correlation past this momentum.  In practice we may have to individually tune $\qmax$ for each term in the spherical harmonic expansion of the correlation.

\begin{figure}
	\includegraphics[width=0.45\textwidth,clip]{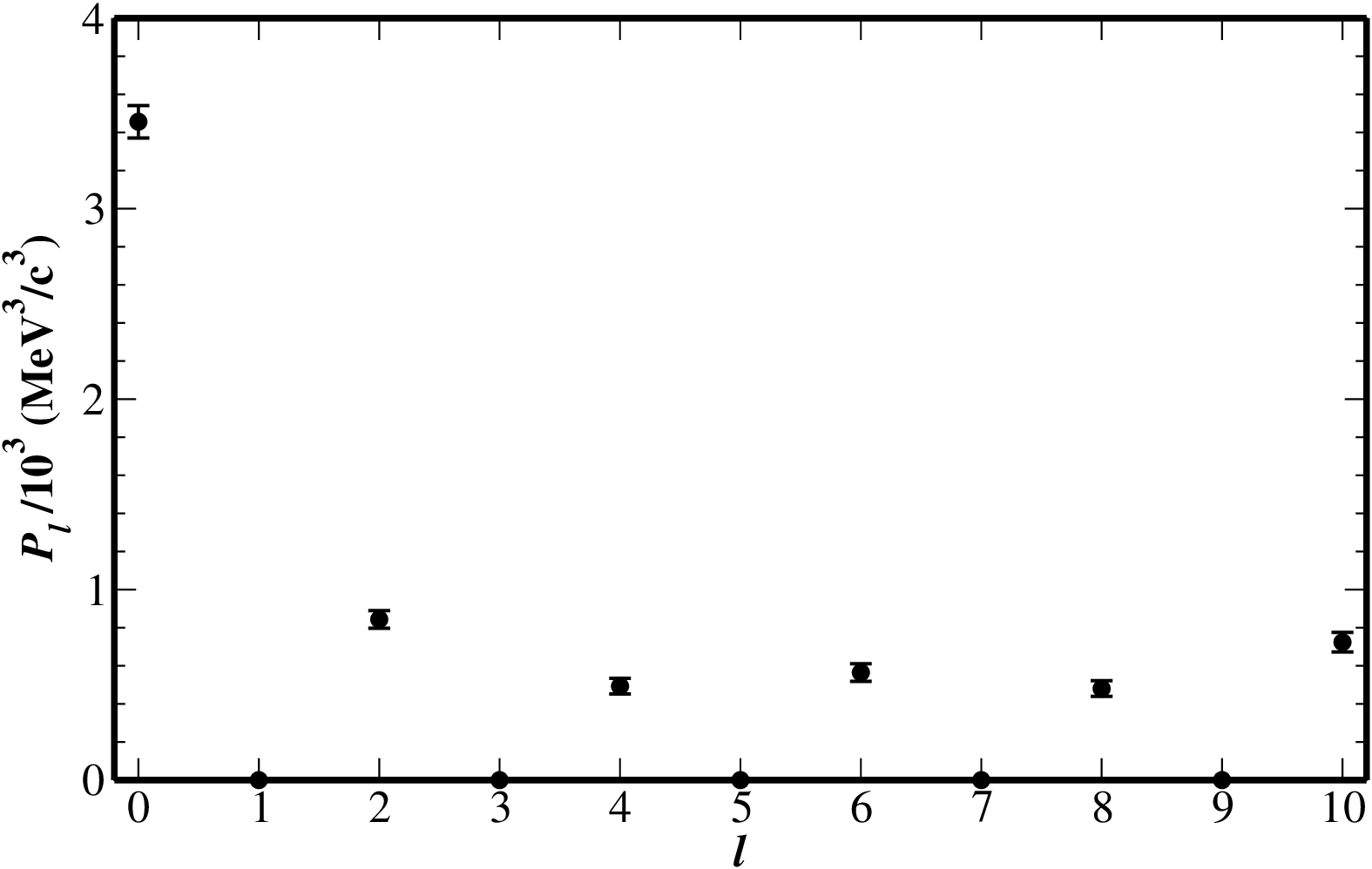}
	\caption{Angular power spectrum of the model correlation function.  }
	\label{fig:power_spec}
\end{figure}

A related measure of the correlation is the spectral power:
\begin{equation}
{\cal P}_{\ell}(\qmax) = 4\pi \sum_{m=-\ell}^\ell\int_0^{\qmax} dq q^2 
	\left|{\cal R}_{\ell m}(q)\right|^2
    \label{eqn:power_spectrum}
\end{equation}
This integral gives an indication as to how much of the correlation is in each $\ell$ in the $Y_{\ell m}$ expansion of the correlation.  The correlation is squared in Eq. \eqref{eqn:power_spectrum}, allowing us to assess the angular content of a correlation even if it dips below 1.  In Fig. \ref{fig:power_spec}, we show the power spectrum of the model correlation. 
It flattens past $\ell=4$ indicating there is no angular information past this order.  In fact, for this test correlation, it is safe to truncate at an even more aggressive $\ell=2$.

\subsection{Representing the Source}

As in Ref.~\cite{imag_1}, we write the radial dependence in of the source functions in terms of basis splines (denoted by $B_j$)~\cite{de_boor}:
\begin{equation}
   S_{\ell m}(r)=\sum_{j=1}^{N_c}S_{j\ell m} B_j(r)
   \label{expand_source_in_Bjs}
\end{equation}
Basis splines are piece-wise polynomials which are commonly used to approximate functions.  Basis splines are a type of spline that has some computational benefits over more commonly used splines such as cubic splines~\cite{de_boor}.  A detailed description of basis splines is contained in \cite{imag_3,de_boor}. 

To a large extent, the quality of any spline approximation of a function is determined by the locations of the knots.  Knots are the points where the  polynomials that make up each spline are pasted together.  If there is some sharp feature in the function to be approximated, then we need a high knot density there to resolve the feature.  On the other hand, if the function is smooth and slowly varying over some range, it makes sense to have very few knots there.  

We would like to engineer a basis spline expansion such that the approximated
source function agrees with the true source function at a set of collocation points ${\bf x} = \left\{r_0, r_1, ..., r_{n-1}\right\}$ on the interval $[\rmin,\rmax]$.  In other words,
\begin{equation}
    S(r_k) = \sum_i S_i B_i(r_k)
\end{equation}
In general, finding the optimal knots ${\bf t} = \left\{t_0, t_1, ..., t_{N_k-1}\right\}$ that minimize the distance between the original function and the approximation in some function space metric is a surprisingly difficult problem.  De Boor \cite{de_boor} presents several schemes for determining these optimal knots and all of these schemes begin with the nearly
optimal choice of
\begin{equation}
    t_i = \left\{ 
        \begin{array}{ll}
            \rmin & \text{for $0\le i \le k-1$}\\
            \displaystyle\frac{r_{i-k-1}+...r_{i-1}}{k} & 
            \text{for $k\le i \le n$}\\
            \rmax& \text{for $n+1\le i \le n+k+1$}
        \end{array}
        \right.
\label{eqn:setknots}
\end{equation}
Because De Boor states that this choice is often close enough to the optimal knots for practical purposes, we adopt these knots for our work.

Our problem is somewhat more complicated in that we want the optimal knots for the {\em convolution} of the fitted function with some other function, namely the kernel of the Koonin-Pratt equation.  In the next subsection, we will describe how we use the Sampling Theorem to find the collocation points.

\subsection{How the kernel determines the collocation points}

The choice of the collocation points for the source and bins for the correlation play a surprisingly important role in setting the quality of the imaging results.  This issue was explored in Ref.~\cite{imag_2,imag_3} leading to the development of the ``optimized discretization'' technique.  This technique can be costly to apply, so we investigate a computationally cheaper alternative approach.  What may be surprising is that the choice of binning in the data also plays a large role in shaping the quality of the inversion.  This problem is known in astronomical imaging \cite{parker,cobe_1,cobe_2,scales_tutorial} and has its root in the Sampling Theorem in Fourier theory.  In this section, we will outline the conclusions for the binning from the Sampling Theorem, but we will defer a detailed study to a future publication. 

In practice, both ${\cal R}_{\ell m}$ and $S_{\ell m}$ are {\em bandwidth limited} meaning that they are effectively zero outside some range.  For ${\cal R}_{\ell m}$, this is easy to see.  After some high $q$, namely $\qscale$,  the data is fluctuating about zero and is statistically indistinguishable from zero. Similarly, $S_{\ell m}$ is cut-off at some high $\rscale$ because most particles are emitted from the reaction zone with finite extent.  For those particles that are produced from resonance decays, the decay rate drops exponentially and at some point the production rate is effectively zero.

These bandwidth limits have important implications that we now work through.   Since each term in the correlation expansion has compact support, we may expand them in terms of spherical Bessel functions, with the requirement that  ${\cal R}_{\ell m}(\qscale)=0$:
\begin{equation}
    {\cal R}_{\ell m}(q) = \sum_{n=0}^{\infty} {\cal R}_{n\ell m} j_\ell\left(
    \frac{q}{\qscale} \alpha_{\ell n} \right)
\end{equation}
Here, $\alpha_{\ell n}$ is the $n^{th}$ zero of the $\ell^{th}$ spherical Bessel function.  Using the Koonin-Pratt equation we can determine the coefficients in this expansion:
\begin{equation}\begin{array}{rl}
    {\cal R}_{n\ell m} =& \displaystyle \frac{8\pi}{ 
    \left( j_{\ell+1}(\alpha_{\ell n})\right)^2}
    \int_0^{\rscale} \dn{}{r} r^2 S_{\ell m}(r)\\ 
    & \displaystyle\times\left( \int_0^1 \dn{}{\eta} \eta^2 
    j_\ell(\eta\alpha_{\ell n}) K_\ell(\eta \qscale,r)\right)
\end{array}\end{equation}
Here $\eta=q/\qscale$.

We can see the meaning of this equation by considering the simple case of  non-interacting spin-0 bosons:
\begin{equation}
    {\cal R}_{n\ell m} = \frac{\pi}{2}(-1)^{\ell/2}\frac{1}{\qscale^3} 
    \frac{1}{\left( j_{\ell+1}(\alpha_{\ell n})\right)^2}
    S_{\ell m}(r_{\ell n}),
    \label{eqn:samplingthm_corr}
\end{equation}
where $r_{\ell n}=\alpha_{\ell n}/2\qscale$.  In other words, the value of the source at a few specific points completely determines the correlation.  Since the source also has compact support:
\begin{equation}
    S_{\ell m} = \sum_{n=0}^\infty S_{n\ell
    m}j_\ell(\frac{r}{\rscale}\alpha_{\ell n})
\end{equation}
Using this, we can show that 
\begin{equation}
    S_{n\ell m} = \frac{(-1)^{\ell/2}}{2\pi \rscale^3 
    \left(j_{\ell+1}(\alpha_{\ell n})\right)^2 } C_{\ell m}
    \left(\frac{\alpha_{\ell n}}{2\rscale}\right)
    \label{eqn:samplingthm_sou}
\end{equation}
In other words, the source itself is completely determined by the correlation at a few points.  Both Eqs. \eqref{eqn:samplingthm_corr} and \eqref{eqn:samplingthm_sou} are statements of the Sampling Theorem.

Why are the points 
\begin{equation}
    r_{\ell n} = \frac{\alpha_{\ell n}}{2\qscale}
    \label{eqn:r_colloc}
\end{equation}
and
\begin{equation}
    q_{\ell n} = \frac{\alpha_{\ell n}}{2\rscale}
    \label{eqn:q_colloc}
\end{equation}
so important?  Well, if we specialize to the $\ell=m=0$ case, we see that the size bins in the $q$-space are related to $\rscale$ (and the the size bins in the $r$-space are related to $\qscale$) in a simple manner owing to the fact that $\alpha_{j 0}=j\pi$:
\begin{equation}
  \begin{array}{rl}
    \Delta q&=\pi/2 \rscale\\
    \Delta r&=\pi/2 \qscale
  \end{array}
  \label{eqn:l=0_colloc}
\end{equation}
In other words, the largest scale in one space determines the smallest resolved scale in the Fourier transform space.  Eqs. \eqref{eqn:r_colloc}--\eqref{eqn:l=0_colloc} pose an interesting chicken and egg problem;  these equations imply that there is an optimal binning for a correlation given its underlying source.

Note that the collocation points will be equally spaced for $\ell=0$, but not for higher $\ell$.  In fact, at low $r$, the higher $\ell$ kernels simply do not have the resolution to determine the source: all spherical bessel functions go to zero as $r\rightarrow 0$ like $j_\ell(r)= r^\ell/\ell !!$ \cite{abram+steg}.  This means that there is a ``dead zone'' at $r=0$ for all $\ell$ except $\ell=0$.  This dead zone occurs in the general case as well.  This tells us that the collocation points should be more spaced out at low $r$.    

In practice, interactions distort the kernels from the simple Fourier kernels.    Since the zeros of the kernel determine the spacing of the collocation points in Eqs.~\eqref{eqn:r_colloc} and~\eqref{eqn:q_colloc} for simple Fourier kernels, we expect that the zeros of more complicated like-pair kernels to play the same role.   For the like-meson kernel with Coulomb turned on, the zeros obtain a $q$ dependence and this is easily taken into account.  For non-identical pairs, there are no zeros in the wavefunctions and our method will fail.  We will investigate alternative strategies for non-identical pairs at a later date.

\subsection{3D Imaging}

By combining Eqs.~\eqref{binnified_c2} and~\eqref{KP_expanded}, we convert the Koonin-Pratt equation into a matrix equation:
\begin{equation}
   \boldsymbol{\cal R}=K\cdot{\bf S}.
\end{equation}
We proceed as in Refs.~\cite{imag_1,imag_2,imag_3,tarantola} and apply Bayesian imaging.  In Bayesian imaging, we use Bayes Theorem to represent the probability of a particular source representing the correlation data~\cite{tarantola}.  This probability distribution is Gaussian, i.e. $\propto e^{-\chi^2}$ so the source that maximizes the probability minimizes the $\chi^2$:
\begin{equation}
   \chi^2=(K\cdot{\bf S}-{\boldsymbol{\cal R}})^T\cdot
      ({\Delta^2 {\cal R}})^{-1}\cdot(K\cdot{\bf S}-{\boldsymbol{\cal R}})
\end{equation}
The source that does this is:
\begin{equation}
   {\bf S}={\Delta^2 S}\cdot K^T\cdot 
      (\Delta^2 {\cal R})^{-1}\cdot{\boldsymbol{\cal R}}
\end{equation}
The covariance matrix of this source is:
\begin{equation}
   {\Delta^2 S} = (K^T\cdot (\Delta^2 {\cal R})^{-1}\cdot K)^{-1}.
\end{equation}
         
In order to stabilize the inversion, we can take advantage equality constraints~\cite{tikhonov}.  An equality constraint is a condition on the vector of source coefficients that has the generic form ${\cal C}\cdot{\bf S}={\bf c}$.  One example of such a constraint is that the source has zero slope at the origin, in which case the $\ell m=00$ component obeys
\begin{equation}
   S'_{00}(r \rightarrow 0)=\sum_{i=1}^{N_M} S_{00i} B_i'(r\rightarrow 0)=0.
\end{equation}
With this, the elements of ${\cal C}$ are ${\cal C}_{00i}=B_i'(r\rightarrow 0)$ and ${\bf c}$ has one component given by $c_{00}=0$.  A larger list of possible equality constraints and their physical origin is tabulated in Table \ref{table:eqcons3D}.
\begin{table*}
    \begin{center}
        \begin{tabular}{c|c|c}
            \hline\hline
            Constraint & Purpose & Functional form \\
            \hline\hline
            \;\parbox{2in}{
                $r=0$ is a maximum of $S({\bf r})$\\(for like pairs only)
            }&
            \;\parbox{2in}{
                \vspace*{\baselineskip}
                Constrain the higher $\ell$ components that are
                not well controlled due to the $r^\ell$ dependence of terms
                in the spherical harmonic expansion.
                \vspace*{\baselineskip}
            }\; &  
            $
               \begin{array}{rl}
                  \displaystyle\frac{\partial S_{\ell m}}{\partial
                     r}(r\rightarrow 0)=0&\forall \ell,m \\
                  \displaystyle S_{\ell m}(r\rightarrow 0)=0 &
                    \forall \ell,m,\; \ell\ne 0
               \end{array}
            $\\
            \hline
            $S({\bf r})=0$ at $r=\rmax$ &
            \parbox{2in}{
                \vspace*{\baselineskip}
                Smooth oscillations in the source at high ${\bf r}$ caused 
                by aliasing of statistical and experimental noise in 
                correlation.
                \vspace*{\baselineskip}
            } &  
            $\displaystyle S_{\ell m}(\rmax)=0 \;\;\forall \ell,m$\\
            \hline
            $S({\bf r})=0$ is flat as $r\rightarrow \rmax$&
            \parbox{2in}{
                \vspace*{\baselineskip}
                Smooth oscillations in the source at high ${\bf r}$ caused 
                by aliasing of statistical and experimental noise in 
                correlation.
                \vspace*{\baselineskip}
            } &  
            $
                \displaystyle \frac{\partial S_{\ell m}}{\partial r}(\rmax)=0
                \;\;\forall \ell,m
            $\\
            \hline\hline
        \end{tabular}
        \caption{Equality constraints on the Basis Spline representation for non-spherically symmetric sources.  Constraints on higher order derivatives are possible, but we use $3^{rd}$ degree splines so cannot apply these higher order derivative constraints.}
        \label{table:eqcons3D}
    \end{center}
\end{table*}

Equality constraints correspond to a specific form of prior information that are easily included in Bayes Theorem~\cite{imag_3,tarantola}.  Adding this prior information amounts to adding a penalty term to the $\chi^2$:
\begin{equation}
   \chi^2+\lambda({\cal C}\cdot{\bf S}-{\bf c})^2
\end{equation}
Here $\lambda$ is a trade-off parameter and we may vary it in order to emphasize stability in the inversion (by making $\lambda$ huge) or to emphasize goodness-of-fit (by setting $\lambda$ to zero).  Such an ability to trade-off stability for goodness-of-fit is discussed in {\em Numerical Recipes} \cite{NumRecipes} in detail.  With this modification of the $\chi^2$, the imaged source is
\begin{equation}
   {\bf S}={\Delta^2 S}\cdot \left(K^T\cdot (\Delta^2 {\cal R})^{-1}\cdot
    {\bf {\cal R}}+\lambda {\cal C}^T\cdot {\bf c}\right),
\end{equation}
and the covariance matrix of source is
\begin{equation}
   {\Delta^2 S} = \left(K^T\cdot (\Delta^2 {\cal R})^{-1}\cdot K
   +\lambda {\cal C}^T\cdot{\cal C} \right)^{-1}.
\end{equation}
An alternative approach is to use Lagrange multipliers to force the constraints to be obeyed.  We have investigated this approach and found the results to be equivalent.

\section{Imaging analysis of test problem}

We now turn to the analysis of the test problem.  We will do this in three stages.  In the first, we will analyze the $\ell m = 00$ term in detail since we can perform cross-checks of this term with the {\tt CorAL} code \cite{coral} and because this term contains several interesting physics elements.  In the second stage, we will summarize the analysis of the higher $\ell$ terms.  In the final stage, we collect all of the terms and present the final results of this test.  

\subsection{The $\ell m=00$ term}

\begin{figure}
    \includegraphics[width=0.45\textwidth,clip]{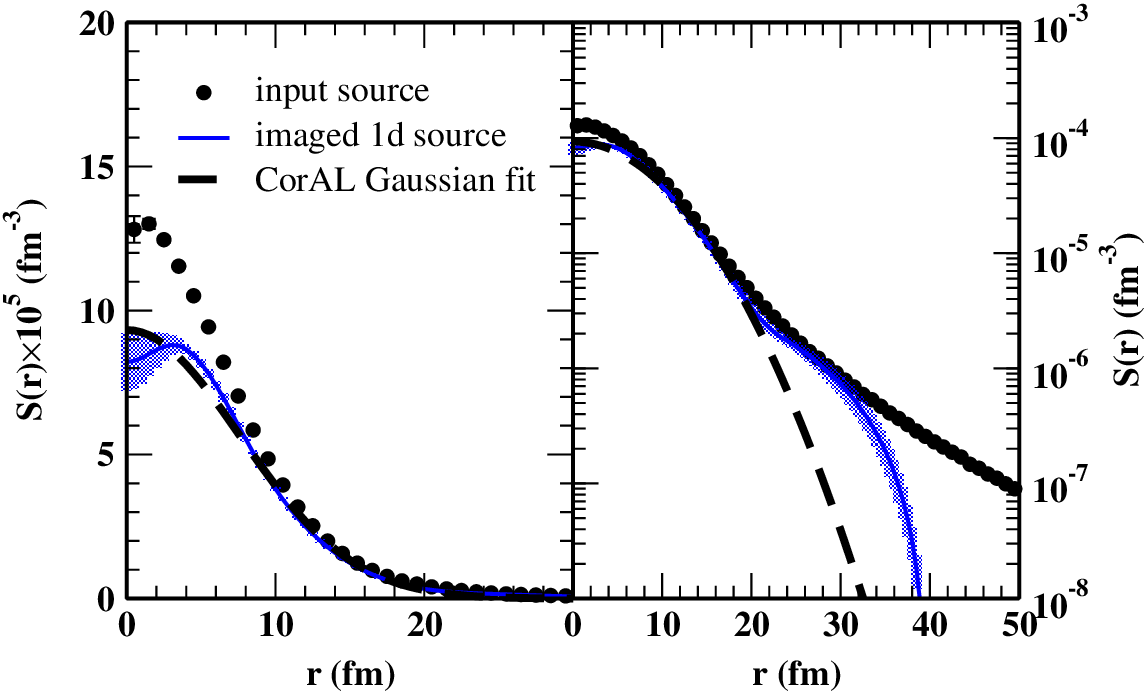}
    \caption{Angle averaged imaged and original sources, compared to best-fit Gaussian source.  On the left we plot them on a linear scale and on the right on a logarithmic scale.}
    \label{1d_fake_sou}
\end{figure}

\begin{figure}
    \includegraphics[width=0.45\textwidth,clip]{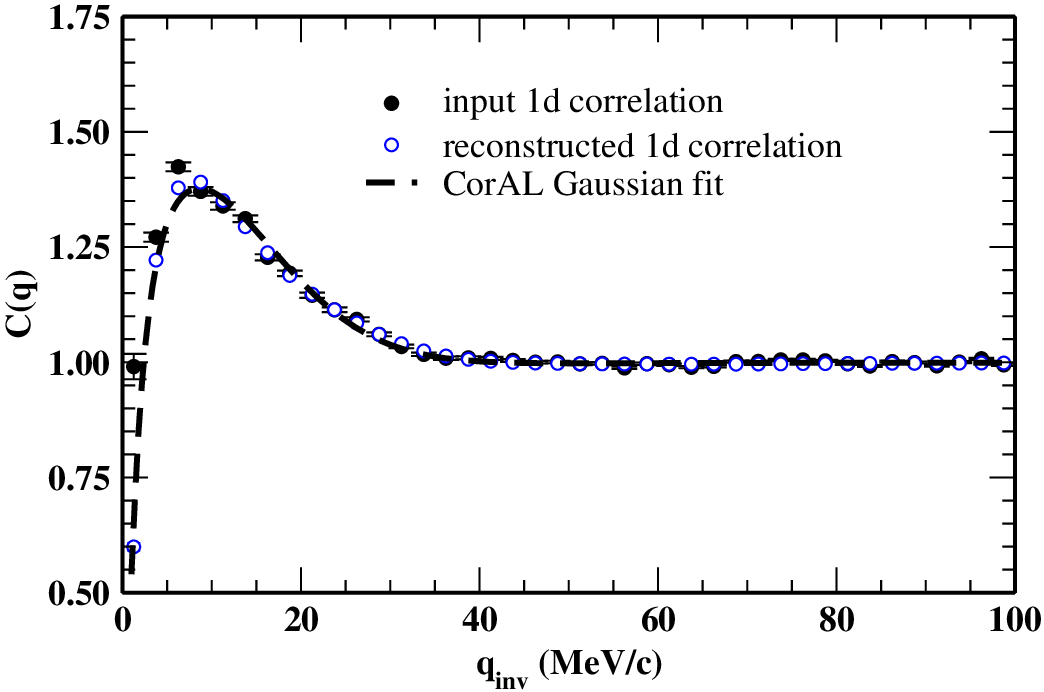}
    \caption{Angle averaged restored correlation, original correlation and correlation from best-fit Gaussian source.}
    \label{1d_fake_corr}
\end{figure}

The $\ell m=00$ term is the angle averaged, $\qinv$, correlation.  Here, we describe the imaging results and compare it to Gaussian fits performed with {\tt CorAL}.  We compare the restored and the fitted sources to the input model source in Fig \ref{1d_fake_sou} and to the input model correlation in Fig \ref{1d_fake_corr}.  In the model source, one can clearly see a non-Gaussian tail caused by the time dependence of the halo term.

We produced the imaging results shown in Figs. \ref{1d_fake_sou} and \ref{1d_fake_corr} using knots set using the Sampling Theorem with $\qmax=39$ MeV/c and 8 3$^{rd}$ degree basis spline coefficients.  We cut the image off at 40 fm and used the constraints in Table \ref{table:eqcons3D} to control the source behavior at the origin and at 40 fm.  We obtained a final $\chi^2/d.o.f. = 186/39$.  In Fig. \ref{1d_fake_sou}, we see that the Gaussian core is well resolved and we can begin to resolve the non-Gaussian tail at least an order of magnitude down from the core.  If we convert the source height and half width to equivalent Gaussian numbers, we find $\lambda = 0.704\pm 0.160$ and $\Rinv= 5.77\pm 0.37$~fm.  Finally, we note the excellent agreement between the original correlation and the correlation produced by uninverting the imaged source.

We also used {\tt CorAL} code \cite{coral} to directly fit a Gaussian source to the model correlation.  {\tt CorAL} works by folding a Gaussian source with the full pion kernel and varying the Gaussian source parameters to maximize agreement with correlation data.  This approach is markedly better than the more commonly used Coulomb corrections used in correlation analysis.  The Gaussian source that {\tt CorAL} uses is:
\begin{equation}
    S(r) = \frac{\lambda}{(2\sqrt{\pi} \Rinv)^{3}} \exp\left(-\frac{r^2}{4\Rinv^2}\right).
\end{equation}
This parameterization is chosen to conform to radii in emission function in Eq.~\eqref{eqn:emission_func} and is consistent with standard radius definitions (i.e. for 1d Gaussian $\pi^0$ correlation, $C(\Qinv)=1+\lambda\exp(-\Qinv^2\Rinv^2)$).  We find best fit values of $\Rinv = 5.380 \pm 0.134$ and $\lambda = 0.647 \pm 0.029$, consistent with image.  We comment that the Gaussian fit to the correlation is slightly worse than that of the image as we see in both the high $q$ fall off and the Coulomb hole in Fig~\ref{1d_fake_corr}.

Comparing both source extraction methods, we see that both the fit and the image reproduce the radii of the input source, but both are approximately 20\% low at the lowest $r$.  In both cases, the sensitivity of of $r^2 K_{\ell}(q,r)$ in Eq. \eqref{KP_expanded} reduces our ability to extract information about the source as $r \rightarrow 0$.
%In both cases, this is a result of the $r^2$ weighting in the kernel integral in the Koonin-Pratt equation (Eq. \eqref{KP_expanded}).  
Improving the input correlation with higher statistics improves the agreement between input and extracted sources but is not always realistic.  Making the data binning smaller can improve agreement with data but the statistics of the data may not support a finer binning.  

Given the excellent agreement between the Gaussian fits and the source images, we wondered it if is possible to extract the fraction of the source in the core directly from the fits or from the equivalent Gaussian parameters from the image.  If the Gaussian is representative of the core in the single particle source, we would expect that $\lambda \approx f^2$ in Eq. \eqref{eqn:emission_func}.  Given the fitted $\lambda$, we find $f=0.804\pm 0.018$.  The fact that $f$ comes out 25\% high is due to non-Gaussianness of the core itself.  The $\Rout$ and $\Rlong$ directions are both strongly distorted because of the time emission in the model and the Gaussian fits are not as good as they appear.  The source integral is a better measure of the source except that it cannot separate out tail and core.  Integrating, we find that the integral is $0.658\pm 0.009$ out to $25$ fm and $0.743\pm 0.015$ out to $40$ fm, meaning 65.8\% of pairs are emitted less than 25 fm apart and 74.3\% or pairs are emitted less than 40 fm apart.

\subsection{The higher $\ell$ terms}

We have imaged the source coefficients up to $\ell =4$.  In all cases, we used the Sampling Theorem knots, tuned the choices of $\rmax$ and $\qscale$ to minimize the $\chi^2$ to the data, and used all the constraints in Table~\ref{table:eqcons3D}.  Our parameters are summarized in Table \ref{table:imaging_parameters}.  

\begin{figure*}
    \includegraphics[width=0.75\textwidth,clip]{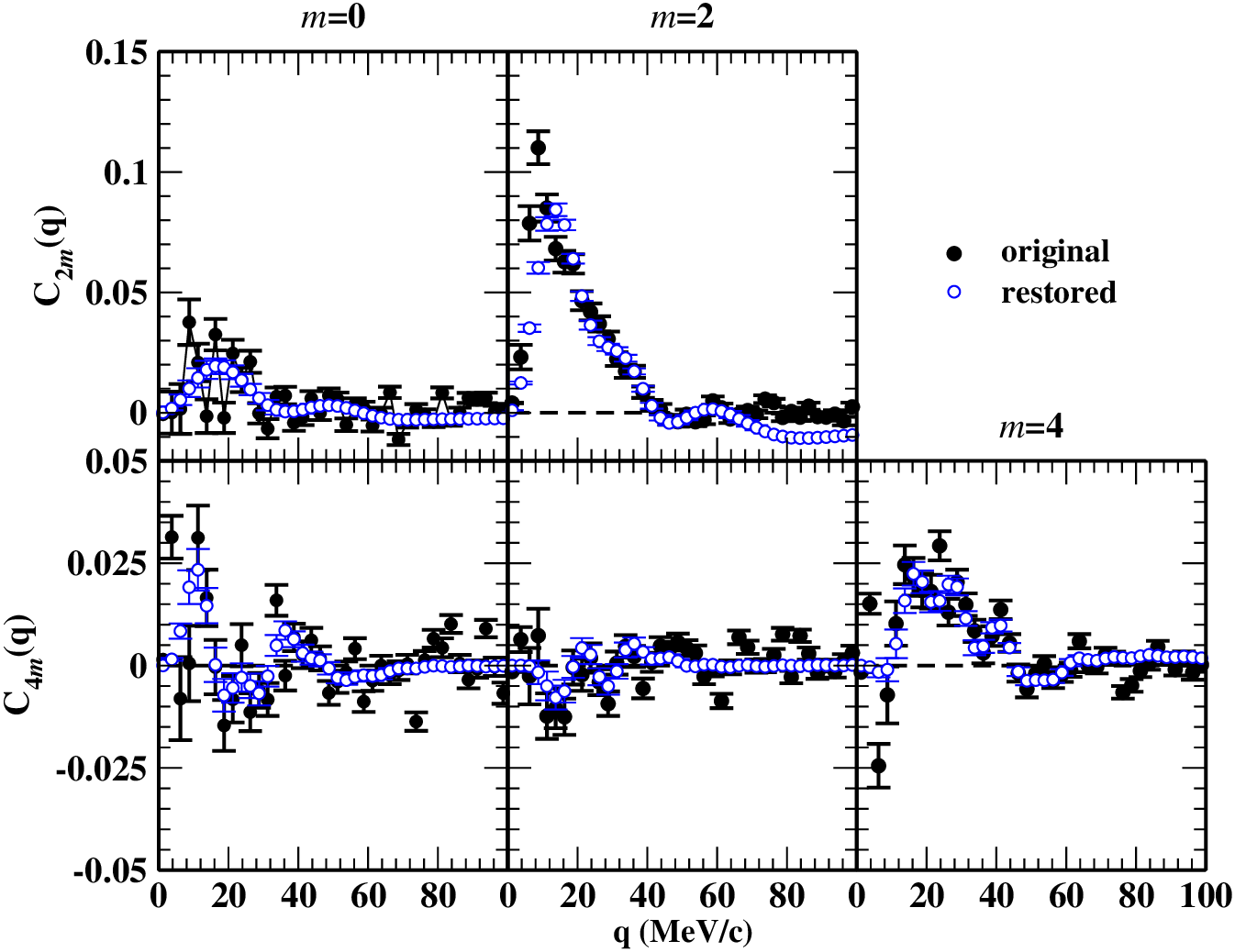}
    \caption{Comparison of the input and restored correlation terms with $\ell > 0$.  The input correlation is shown in black symbols and the restored source in the open symbols.}
    \label{fig:higher_l_fake_corr}
\end{figure*}

Together these terms do not modify the $r=0$ fm behavior of the full 3d source because of the leading $r^{2+\ell}$ behavior of kernel and integration measure.  They do modify the large $r$ fall-offs at larger r and hence the source radii.  We show the restored correlations in Fig.~\ref{fig:higher_l_fake_corr}.  They are not exceptionally enlightening but do show the overall good agreement below $\qmax$.  We do not use data points with $q>\qmax$, so any agreement there is fortuitous but not required.  However, we note the rather high $\chi^2$ and poor fit to the $\ell m=22$ term.  We attribute this to the non-Gaussian in the tail in the outwards direction causing the relatively sharp rise in this term of the correlation, which we have difficulty resolving.  We have not expanded the test source in $Y_{\ell m}$'s so do not show the source term comparison.

\begin{table}
\begin{tabular}{cc|c|c|c|c}
\hline\hline $\ell$ & $m$ & \parbox{.6in}{\vspace{1mm}$\qscale$\\(MeV/c)\vspace{1mm}} & \parbox{.3in}{$\rmax$\\(fm)} & \parbox{.5in}{$\qmax$\\(MeV/c)} & $\chi^2/d.o.f.$ \\\hline \hline
0 & 0 & 39   & 40 & 60 & 186/39 \\ \hline
2 & 0 & 63.5 & 40 & 60 & 189/38 \\ \hline
2 & 2 & 77   & 40 & 60 & 665/38 \\ \hline
4 & 0 & 70   & 60 & 60 & 187/38 \\ \hline
4 & 2 & 50   & 60 & 60 & 162/38 \\ \hline
4 & 4 & 70   & 60 & 60 & 174/38 \\ \hline\hline
\end{tabular}
\caption{Parameters used in the imaging analysis of the test problem.}
\label{table:imaging_parameters}
\end{table}

\subsection{Collecting the terms: full 3d correlation and source}

\begin{figure}
	\includegraphics[width=0.45\textwidth]{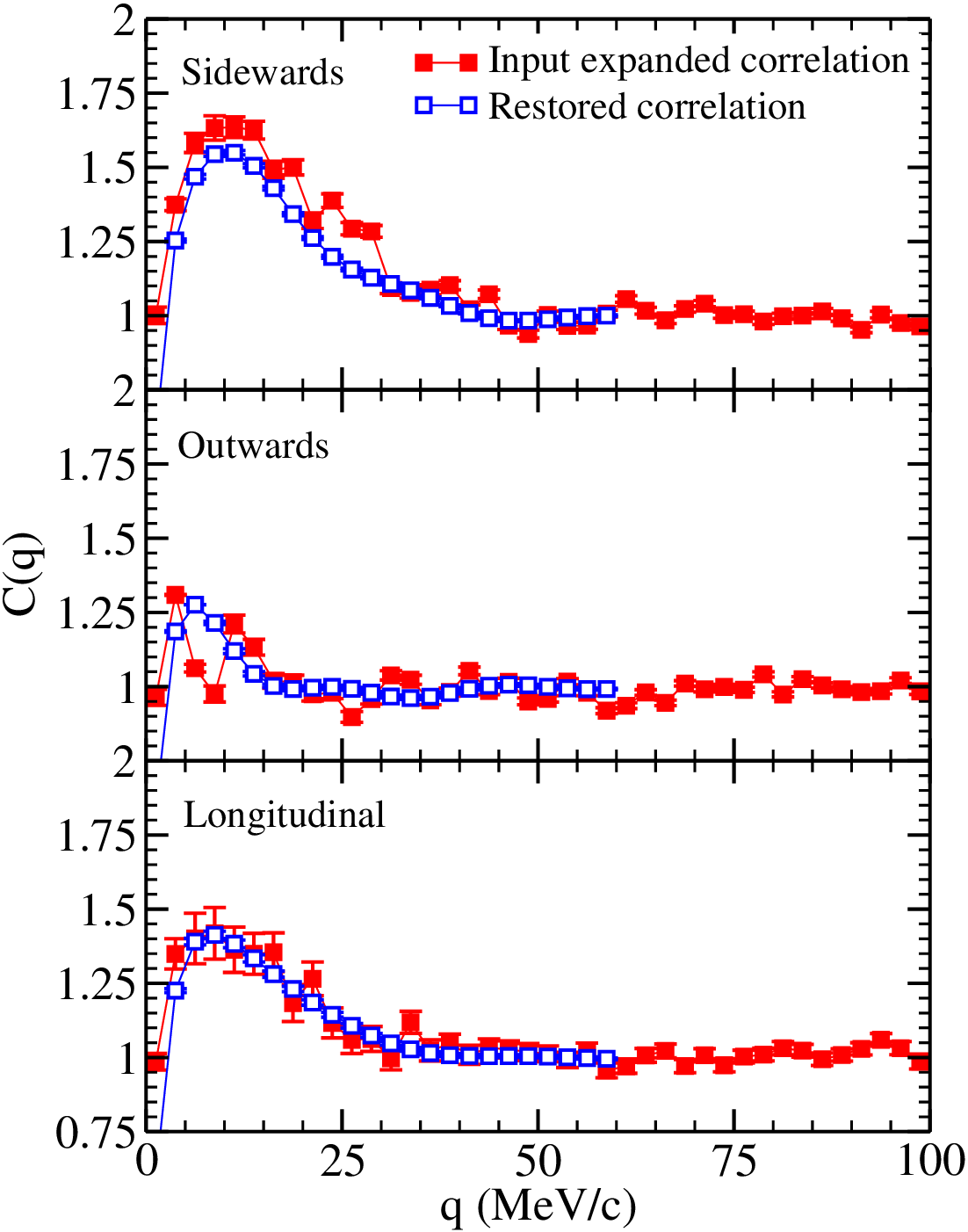}
	\caption{The input model correlation compared with the correlation expanded in spherical harmonics, stopping at order $\lmax=4$.}
	\label{fig:corr_compare_fakedata}
\end{figure}

We now collect the source and correlation terms and compare them to the full 3d input data.  We show the correlation in Fig~\ref{fig:corr_compare_fakedata}.  The restored correlation is consistent with input, but is significantly smoother.  The agreement between them is excellent for the higher $\ell$ terms and not bad for the 00 term.  We reproduce both the Coulomb hole and the large $q$ fall-off.

\begin{figure}
   \includegraphics[width=3.25in]{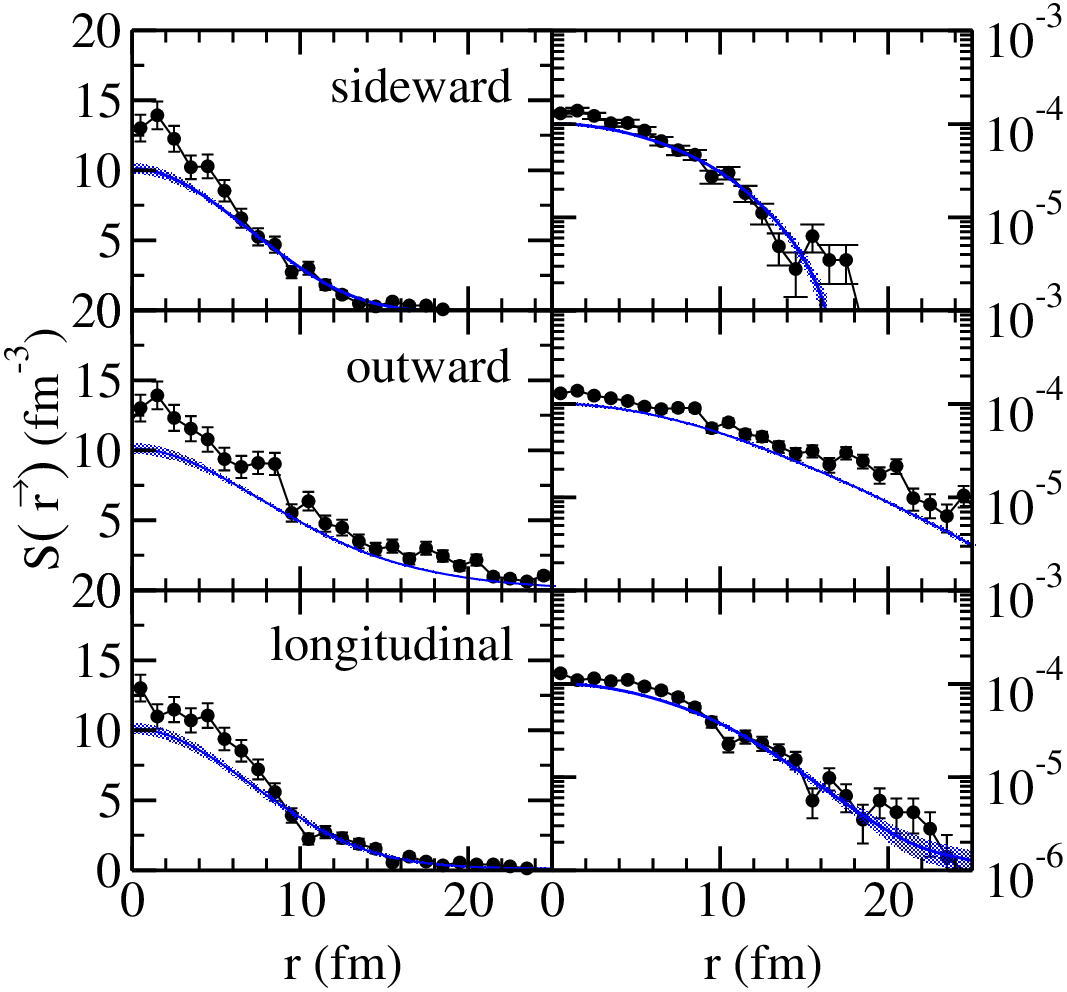}
   \caption{The model source function constructed from the model emission functions in Eq.~\eqref{eqn:emission_func}.  The imaged source is the curve with the grey error band.  We produced the imaged source using only the $r=0$ constraints and we truncated the expansion at $\lmax=4$.}
   \label{fig:sou_compare_fakedata}
\end{figure}

Finally we turn the imaged source shown in Fig~\ref{fig:sou_compare_fakedata}.  Looking first in the sidewards direction, we clearly see the Gaussian core alone.  In the outwards and longitudinal directions, we also clearly see the non-Gaussian tails caused by the time dependence of the halo term in Eq \eqref{eqn:emission_func}.  As in the $\qinv$ correlation, we have the same 20\% short fall at low $r$ because of the $\ell,m=00$ term.  Reading off the half widths and source height, we can compute equivalent Gaussian radii.  We find $\Rout = 7.02\pm 0.25$, $\Rside = 3.42\pm 0.21$, $\Rlong = 5.04\pm 0.31$ and $\lambda = 0.442\pm 0.067$.  Interestingly, $\lambda$ surprisingly close to the $f^2$ we put in to the test source.  The radii are a bit higher than the radii we put into the single particle source in Eq. \eqref{eqn:emission_func}.  This is a result of the mixing of the space and time directions in the construction of the source function in Eq. \eqref{eqn:sou-def}.  In the end, it is clear that we not only can image reliably in 3d, but we can do it with the same precision as we could in one dimension.

%-------------------------------------------
\section{Conclusion}
%-------------------------------------------

In this paper, we have introduced an extension to the source imaging method from Refs. \cite{imag_1,imag_2,imag_3} for imaging full three-dimensional correlations.  This extension relies on spherical harmonic expansion of the correlation and source functions, effectively converting an intractable three-dimensional inversion to a set of simple one-dimensional inversions.  We have demonstrated this technique on a seemingly simple model in which a non-Gaussian and non-spherical source tail is generated from a simple Gaussian single particle source with finite lifetime.  Within this model, we have shown that we can resolve the core radii to a level comparable to Gaussian fits.  We have also shown that we go far beyond this by resolving the detailed aspherical shape of any non-Gaussian tails of the source.  

While we have shown the techniques efficacy only for like-pion correlations, imaging is a general tool and is limited only by the availability of data and our ability to compute relative wavefunctions.  Imaging provides a means to measure source shapes and sizes for complicated final states where a Gaussian fit is impossible, such as for non-identical correlations.  Finally, this extension to the imaging technique should shed light on the non-Gaussian tails now being seen in advanced analyses of RHIC two-particle correlations \cite{roypaper}.     
            
%-------------------------------------------
\section*{Acknowledgements}
%-------------------------------------------
We gratefully acknowledge stimulating discussions with Drs. G.~Verde, S.~Panitkin, N.~Xu, K.~Hagino, and S.~Johnson.  

This research is supported by U.S. Department of Energy Grant No. DOE-ER-41132 and National Science Foundation Grant No. PHY-0245009.  This work was performed under the auspices of the U.S. Department of Energy by Lawrence Livermore National  Laboratory under Contract W-7405-Eng-48.
\nobreak

%-------------------------------------------

%=============================================================================
%  "EndMatter"
%=============================================================================

\end{document}